\documentclass{report}
\usepackage{graphicx}
\usepackage{epsfig}
\input{epsf}

\begin{document}

\baselineskip=18pt

\noindent
\textbf{\Large Raman Measurements and Stress Analysis in
Gallium Ion Implanted Gallium Nitride Epitaxial Layers on
Sapphire}

\vspace*{0.1in} \noindent {\large S. Mal$^a$, A. Singha$^a$, S. Dhara$^b$, A. Roy$^{a*}$}\\
\emph{$^{a}$Department of Physics, Indian Institute of Technology
Kharagpur 721302, West Bengal, India} \\
$^b${\emph{Institute of Atomic and Molecular Sciences, Academia
Sinica, Taipei 106, Taiwan}}

\baselineskip=32pt
\vspace*{0.5in} \noindent
{\bf Abstract}\\
 In this article,
we estimate hydrostatic stress developed in gallium ion implanted
gallium nitride epitaxial layers using Raman measurements. We have
calculated deformation potential constants for $E_2$(high) mode in
these epi-layers. The presence of a polar phonon-plasmon coupling
in these systems has also been demonstrated. In as-implanted
samples, with an increase in implantation fluence, we have
observed disorder-activated Raman scattering.


\newpage
\noindent {\bf \large 1. Introduction}

Wurtzite structure Gallium Nitride ($h$-GaN) epitaxial layers
(epi-layers) are direct wide band gap ($\sim$ 3.4 eV) two
dimensional semiconductor systems which have applications in
optoelectronic and microelectronic devices operating in the blue
and ultraviolet. In GaN-based devices, ion-implantation is an
attractive technique for selective-area doping, precise control
over dopant concentration etc. Above a threshold fluence,
ion-implantation leads to formation of cubic ($c$-GaN, space group
T$^{2}_{d}$) phase, which has many advantages over $h$-GaN (space
group C$^{4}_{6v}$) \cite{Limmer:1998}. The ion irradiation knocks
out atoms from the irradiated material and creates point defects
in the lattice. It does not create any massive damage and large
defect complexes. It is to be noted that the important point
defects, caused by ion-irradiation, are vacancies and
interstitials. These defects induce localized electron energy
levels into the band gap of the host semiconductor and hence
results in a change in the electrical properties of the material.
In addition, these defect states interact with light, inducing an
increase in absorption or emission of photons in  radiative
recombination processes \cite{Perlin:1995,Neugebauer:1996,
Mattila:1997,Glaser:1995}. It has been claimed that the atomic
structure of the defect states are responsible for the yellow
emission in GaN \cite{Chang:2001}. Strain in the lattice (due to
lattice mismatch, difference in thermal coefficient, incorporated
impurities, and point defects) is an important issue in
fabricating optoelectronic devices.

Stress-strain relation in virgin $h$-GaN layer is well-studied in
the literature \cite{Dav:1997}. In this article, we discuss Raman
measurements on self-ion (Ga$^{++}$) implanted GaN epi-layers,
with an emphasis on induced stress in the system. From our
experimental results, we have estimated the hydrostatic strain
coefficients inside the layers. In addition, we show the effect of
disorder-activated Raman scattering in as-implanted samples and
demonstrate the presence of a  polar phonon-plasmon coupling in
post-annealed implanted epi-layers.

\noindent {\bf \large 2. Experimental Details}

GaN epi-layers (pristine sample), of thickness 6 $\mu$m, were
grown on (0001) Al$_2$O$_3$ substrate in a horizontal Metalorganic
Chemical Vapor Deposition reactor at atmospheric pressure.
Trimethylgallium (TMGa) was used as precursor with NH$_3$ as
reactant gas. The flow rate of NH$_3$ was 3000 sccm. The
low-temperature-deposited GaN buffer layer was first grown at
550°C for about 30 min. Temperature was raised to 1050°C and
growth is continued for 8-10 hrs. Ga$^{++}$ ions were implanted in
these layers at 3.0 MeV with different fluences, $1\times10^{15}$,
$2\times10^{15}$ and $1\times10^{16}$ per cm$^2$ (Sample A -
Sample C). The damage of the crystal lattice, induced by
implantation, was removed by thermal annealing in flowing
ultra-high pure N$_2$ at 650 $^\circ$C for 15 minutes and
subsequently at 1000 $^\circ$C for 2 minutes \cite{Liu:2001}. The
undoped and implanted epi-layers were found to be of n-type from
Hall measurements with carrier concentration $\sim$ 4-8 $\times$
10$^{17}$ cm$^{-3}$. Raman measurements were carried out at room
temperature in a back-scattering geometry using a 488 nm air
cooled Ar${^+}$ laser as an excitation source. To avoid phase
change in the samples by laser heating, the power density on the
samples was tuned to 3 $\times$ 10$^4$ Watt/m$^2$. The Raman
spectra were obtained using TRIAX550 single monochromator with an
open electrode charge coupled device as a detector. With 150
$\mu$m slit-width of the spectrometer, the accuracy of our Raman
measurement is $\pm$1.5 cm$^{-1}$.

\noindent {\bf \large 3. Results and Discussion}

\noindent {\bf 3.1. Pristine and self-ion implanted GaN epi-layer}

Fig. 1 shows the Raman spectra of pristine and as-implanted GaN
layers over a wavenumber range of 240 to 850 cm$^{-1}$.


\noindent {\bf Pristine GaN epi-layer:}

Group theory predicts that for $h$-GaN there are eight phonon
modes, $2A_1+2B_1+2E_1+2E_2$, at the $\Gamma$ point.  Except one
$A_1$ and one $E_1$ mode (which are acoustic), the remaining six
modes are optical. Optical $A_1$ and $E_1$ modes are both Raman
and infrared active, while the two $E_2$ modes are only Raman
active and the two $B_1$ modes are silent (Raman inactive). The
higher frequency branch of the $E_{2}$ phonon mode is denoted by
$E_{2}^{H}$ \cite{Harima:2002}. Peaks at 374, 416 and 751
cm$^{-1}$, shown by $\ast$ marks in Fig. 1, are phonons from the
sapphire substrate \cite{Porto:1967}. In the Raman spectrum of the
pristine sample the line at 568 cm$^{-1}$ is due to the
$E_{2}^{H}$ mode. The Raman mode at 345 cm$^{-1}$ is not allowed
by the C$_{6v}$ space group in the first order Raman scattering at
the zone center. This peak can be attributed to the 2nd order
Raman scattering due to acoustic overtones in this region (300 -
420 cm$^{-1}$) \cite{Siegle:1997}. In the high frequency region
for the pristine sample, we distinctly observe a feature at 743
cm$^{-1}$ (shown by double arrows in Fig. 1) due to the polar
mode. This appears as a shoulder of the prominent peak at 751
cm$^{-1}$ from sapphire. The origin of this peak will be discussed
latter.

\vspace{0.5in}
\noindent{\bf Self-ion implanted GaN epi-layers:}\\
After
ion-implantation
 the system loses its crystallinity and becomes disordered.
This is reflected in the broadening of the spectra of
ion-implanted GaN layers (Fig. 1). In addition, for as-implanted
samples, we observe a broad band at around 290 cm$^{-1}$, which
becomes prominent at higher fluences (inset of Fig. 1). As a
result of ion- irradiation, there is a slight rearrangement in the
lattice structure, hence, the wavevector conservation rule in
Raman scattering gets relaxed. The Raman spectrum reflects the
total phonon density of states (so called disorder-activated Raman
scattering). The Raman-inactive modes become Raman-active. Though
silent for bulk $h$-GaN, the phonon density of states of the
highest acoustic phonon branch at the zone boundary, the
$B_{1}^{L}$ mode, has a strong feature in the region 290-300
cm$^{-1}$ \cite{Bungaro:2000,Srivastava:2000,Karch:1997}. We
assign the above observed feature at around 290 cm$^{-1}$ to this
$B_{1}^{L}$ mode \cite{Limmer:1998}. Due to disorder-activated
Raman scattering, the $B_{1}^{H}$ mode also becomes allowed and
appears as a small hump around 674 cm$^{-1}$, shown by $\circ$
mark in Fig. 1 (See phonon density of states in Ref.
\cite{Karch:1997}). The polar mode at around 740 cm$^{-1}$ is also
broadened. In these as-implanted samples, the absence of Raman
lines from sapphire (substrate) indicates the effective
``trapping" of the laser light inside the layers, which can affect
the Raman scattering from the layers by changing the selection
rules.


\noindent {\bf \large 3.2. Post-annealed implanted samples}\\
Raman spectra of the post-annealed GaN layers are shown in Fig. 2.
After annealing, the disorder in samples gets reduced. All Raman
modes in the low frequency region, discussed for the pristine
sample, reappear distinctly in their spectra. In Raman spectra of
post-annealed implanted samples the line at 573 cm$^{-1}$ is due
to $E_{2}^{H}$ mode, shifted by 5 cm$^{-1}$ from its value (568
cm$^{-1}$) in bulk \cite{Harima:2002}. However, we do not observe
any variation in spectral line-shape and position with
implantation (Sample A - Sample C). Peaks from the sapphire
substrate clearly reappear in these spectra. With the decrease in
disorder in the annealed samples, the intensity of the peak at 290
cm$^{-1}$ due to the disorder-activated Raman scattering also
decreases (inset of Fig. 2). However, we would like to point out
that for $c$-GaN LA(X) and LA(L) (LA: longitudinal acoustic) modes
appear at 286 cm$^{-1}$ and 296 cm$^{-1}$ \cite{Zi:1996}. The
photoluminescence spectra of our samples (not shown here) do not
carry a clear signature of the presence of cubic phase in them.


\noindent {\bf \large 3.3 LO phonon -plasmon coupling}\\ In Fig. 3
we have shown Raman spectra in [z(y,y)\={z}] and [z(x,y)\={z}]
configurations for Sample A. The low frequency acoustic mode at
345 cm$^{-1}$ and high frequency polar mode at 743 cm$^{-1}$
appear only in parallel polarization. Similar polarization
dependence of the polar modes has been observed for all
post-annealed implanted samples. In $h$-GaN, polar $A_{1}$(LO)
[LO:longitudinal optical] mode is expected to appear at 733
cm$^{-1}$ in [z(y,y)\={z}] configuration only. Assigning the
observed feature at 743 cm$^{-1}$ to this $A_{1}$(LO) mode is not
justified, because the shift of a Raman line by 10 cm$^{-1}$
cannot be explained either by strain or zone folding effect
\cite{Li:2002}. Also, it is to be noted that this is the only peak
for which the frequency shift is more than the shift for the other
peaks in Fig. 2.

Undoped or nominally doped GaN is invariably of n-type, usually
with a high free electron concentration (10$^{17}$ -10$^{18}$
cm$^{-3}$) at room temperature \cite{Madelung}. The coupling of
collective electronic excitation (plasmon) and lattice vibration
gives rise to the LO phonon-plasmon coupled mode. Such a mode in
$h$-GaN shifts the LO phonon to the higher frequency
\cite{Harima:2002}. Assigning the Raman line at 743 cm$^{-1}$ to
the plasmon-A$_{1}$(LO) phonon mode, we estimate the charge
carrier density in Sample A using the empirical relation
\cite{Harima:2002}
\begin{equation}
n=1.1 \times 10^{17} (\Delta\omega)^{0.764},
\end{equation}
where, $\Delta\omega$ is the observed frequency shift of the LO
phonon. Taking $\Delta\omega$ = 10 cm$^{-1}$, the $n$ is estimated
to be 6 $\times$ 10$^{17}$ cm$^{-3}$. which is also the value
obtained from the Hall measurement. The above equation is valid
for $n \leq 1 \times 10^{19}$ cm$^{-3}$.

Considering the contribution of the deformation potential within
the lattice, a more precise evaluation of $n$ can been carried
out. The intensity of the Raman scattering  is then given by
\cite{Harima:2002,Klein:1972,Irmer:1983}
\begin{equation}
I(\omega)=SA(\omega){\mbox Im}[-1/\epsilon(\omega)],
\end{equation}
where, $\omega$ is the Raman shift and $S$ is the proportionality
constant. The dielectric function is
\begin{equation}
\epsilon(\omega)=\epsilon_{\infty}[1+(\omega^2_L-\omega^2_T)/(\omega^2_T-\omega^2-i\omega\Gamma)-\omega^2_P/(\omega^2-i\omega\gamma)]=0.
\end{equation}
Here, $\omega_T$ and $\omega_L$ are the transverse optical (TO)
and LO phonon frequencies for the uncoupled phonon modes.
$\epsilon_{\infty}$ is the high-frequency dielectric constant of
GaN. $\Gamma(\gamma)$ is the damping rate of the phonon (plasmon).
The plasmon frequency $\omega_P$ and $A(\omega)$ are given by
\begin{equation}
\omega_p=[4\pi ne^2/(\epsilon_\infty m^\star)]^{1/2}
\end{equation}
and
\begin{eqnarray}
A(\omega)=1+2C\omega^2_T[\omega^2_p\gamma(\omega^2_T-\omega^2)-\omega^2\Gamma(\omega^2+\gamma^2-\omega^2_P)]/\Delta\omega\nonumber\\
+C^2(\omega^4_T/\Delta\omega)[\omega^2_P[\gamma(\omega^2_L-\omega^2_T)
+\Gamma(\omega^2_P-2\omega^2)]+\omega^2\Gamma(\omega^2
+\gamma^2)]/(\omega^2_L-\omega^2_T).
\end{eqnarray}
Here, $C$ is the Faust-Henry coefficient, can be estimated from
the ratio of the intensities of the polar LO and TO Raman
scattering modes measured at 90$^\circ$ geometry
\cite{Faust:1966}. The effective mass of the electron is
$m^{\star}$ = 0.2$m_{0}$. All three parameters, $\Gamma$, $\gamma$
and $n$, are free parameters, which need to be chosen correctly,
by fitting more than one Raman spectra of the layers with
different carrier concentrations. Unfortunately, the Raman line
shapes of all our post-annealed samples are nearly same, except,
with higher implantation dose, a slight decrease in intensity of
the coupled phonon-plasmon mode (modified $A_{1}(LO)$ mode) is
observed due to damping by the hole plasmon. In such a case, it
was not possible to choose all three parameters , $\Gamma$,
$\gamma$ and $n$ meaningfully, to estimate more accurate value of
$n$ by following this procedure, compared to what we have obtained
from  Eqn. 1.

\noindent {\bf \large 3.4 Stress Analysis}

An epi-layer, of thickness 6 $\mu$m, is not expected to exhibit a
shift in the Raman line due to the effect of confinement of
phonons \cite{campbell:1986}. Thus, the origin of the shift in
Raman frequency from bulk samples can be either due to  the
presence of elastic strain or due to the change in interaction
between the elastic medium and the macroscopic field. To eliminate
the second factor we have investigated the non-polar $E_{2}^{H}$
mode. It is important to note that within our experimental
accuracy,  all annealed Ga$^{++}$ implanted GaN epi-layer exhibit
nearly identical Raman spectra. Thus, the following stress
analysis holds good for  all post-annealed implanted samples.

GaN epi-layers grown on sapphire contain residual strains produced
by mismatch in the lattice constants and thermal expansion
coefficients between the GaN film and substrate
\cite{Strite:1992}. Sufficient residual strain in the film induces
formation of dislocations and stacking faults. Residual strain
also leads to wafer bowing. It has been shown in Ref.
\cite{Park:2005} that for the above defects in a GaN layer of
thickness 1 $\mu$m (on sapphire substrate) the bi-axial strain
energy varies from 0 (GaN surface) to 0.4 GPa (GaN/sapphire
interface), indicating that the surface layer is free from this
type of strain. On the other hand, the ion irradiation induces
strain in the film due to point defects. These defects are
expected to be more near the surface of the layer. As the
thickness of our films is 6 $\mu$m, we attribute the shift in
non-polar Raman mode only to the strain, generated by the point
defects in the layer. These defects induce three dimensional
stress, such as hydrostatic stress, if they are uniformly
distributed. Based on the assumption of pure elastic theory (where
Hooke's law is valid) the compressive stress can be calculated
from the shift in $E_{2}^{H}$ from \cite{gleize:2003}
\begin{equation}
\omega-\omega_0 = 4.17\sigma_{H},
\end{equation}
where, $\sigma_{H}$ is the hydrostatic stress in GPa, $\omega$ and
$\omega_{0}$ are the Raman line position of the stressed and
unstressed samples. For a shift of 5 cm$^{-1}$ of the $E_{2}^{H}$
mode between bulk GaN and ion-implanted layers we have estimated
$\sigma_{H}$ $\approx$ 1.20 GPa (see Table I).

Knowing $\sigma_{H}$, the hydrostatic strain tensor components,
$u_{xx}$ and $u_{zz}$ (inplane and normal components of strain
tensor) can be estimated from the following relations
\cite{landau}
\begin{equation}
u_{xx} = u_{yy} = \sigma_{H}/Y
\end{equation}
and
\begin{equation} u_{zz} = -R u_{xx}.
\end{equation}

\noindent The Young's modulus, $Y$, of the material, in terms of
elastic stiffness constants $C_{ij}$ is given by
\cite{gleize:2003}
\begin{equation}
Y = \frac{(C_{11}+C_{12})C_{33}-2C_{13}^{2}}{C_{33}-C_{13}}
\end{equation}

\noindent
 and $R$ is the Poisson's ratio, can be written as

\begin{equation}
R  = \frac{C_{11}+C_{12}-2C_{13}}{C_{33}-C_{13}}.
\end{equation}

\noindent Using Eqn. 7 to Eqn. 10, the strain tensor components
have been estimated and tabulated in Table I. Here, the elastic
constants $C_{ij}$ are taken to be $C_{11}$ = 396, $C_{12}$ = 144,
$C_{13}$ = 100, and $C_{33}$ = 392 GPa \cite{Kim:1996}.

The change in frequency for a given phonon mode under symmetry
conserving stress can be expressed in terms of strain tensor
components and deformation potentials constants ($a_{\lambda}$ and
$b_{\lambda}$)  as \cite{gleize:2003}
\begin{equation}
\omega - \omega_{0} = 2a_{\lambda}u_{xx} + b_{\lambda}u_{zz}.
\end{equation}

Furthermore, the bulk Gr\"{u}eneisen parameter $\gamma_{\lambda}$
is related to the characteristic phonon frequency at hydrostatic
compression, phonon deformation potential constants, and elastic
constants of $h$-GaN by \cite{Dav:1997}

\begin{equation}
\gamma_{\lambda}=-\frac{2a_{\lambda}(C_{33}-C_{13})+b_{\lambda}(C_{11}+C_{12}-2C_{13})}{\omega_{\lambda}
(C_{11}+C_{12}+2C_{33}-4C_{13})}.
\end{equation}

\noindent For GaN, $C_{33}+C_{13}=C_{11}+C_{12}$. Therefore, from
the above relation we get,
\begin{equation}
\gamma_{\lambda}=-\frac{2a_{\lambda}+b_{\lambda}}{3\omega_{\lambda}}.
\end{equation}
The Gr\"{u}eneisen parameter used is $\gamma_{E^{H}_{2}}$ = 1.54
\cite{Dav:1997}. Knowing the values of strain tensor components,
the deformation potential constants have been calculated to be -81
cm$^{-1}$ and -2486 cm$^{-1}$.


 With irradiation,
we do not observe any noticeable change in the peak position and
lineshape of non-polar Raman modes, which implies that no extra
stress has been introduced in layers with increase in implantation
fluence. The values of above parameters $u_{xx}$, $u_{zz}$,
$a_{\lambda}$ and $b_{\lambda}$ are nearly same for all implanted
samples.

\noindent {\bf \large 3.5 Conclusion}\\
 We have performed first-order Raman measurements on
Ga$^{++}$ implanted GaN epi-layers. Raman lines become broadened
after implantation. However, they reappear in all post-annealed
implanted samples. We do not observe any change in peak positions
in Raman spectra of the films  with an increase in implantation
dose. Thus, the estimated hydrostatic stress and strain tensor
components, obtained by us from the knowledge of phonon modes, are
expected to hold good for all post-annealed implanted samples. It
would have been interesting to check the stress analysis by high
resolution x-ray diffraction measurements. We have also the
disorder-activated Raman scattering in as-implanted samples.
Interesting behavior of polar modes is reported for the
post-annealed implanted samples.

\vspace*{1.5cm}

\noindent
{\bf Acknowledgements}

Authors thank Department of Science and Technology, India for
financial assistance. SM also thanks Board of Research in Nuclear
Science, India for financial support. SD thanks Y.C. Yu, Institute
of Physics, Academia, Sinica. Taipei, Taiwan, for his help and
also NSC, Taiwan, for financial assistance. Authors also thank the
reviewers for their suggestions and comments.

\noindent $^*$E-mail:anushree@phy.iitkgp.ernet.in\\ $^\dag$On
leave from Materials Science Division, Indira Gandhi for Atomic
Research, Kalpakkam 603102, India. Email: dhara@igcar.ernet.in.

\newpage

\noindent
{\bf Figure Captions}

\vspace{0.15in} \noindent Figure 1. Raman spectra of pristine (P)
and as-implanted GaN epi-layers (Sample A - Sample C). The region
between 240-340 cm$^{-1}$ is shown in the inset.

\vspace{0.15in} \noindent Figure 2. Raman spectra of pristine and
post- annealed Ga$^{++}$ implanted GaN  epi-layers (Sample A-
Sample C). The region between 240-340 cm$^{-1}$ is shown in the
inset.

\vspace{0.15in} \noindent Figure 3. Polarization dependence of
Raman modes in Sample A. Other implanted samples exhibit similar
behavior.

\newpage

\noindent {\bf Table Caption}

\vspace{0.15in} \noindent Table I. Hydrostatic stress, phonon
deformation potential constants and strain tensor components of
postannealed Ga$^{++}$ implanted GaN epi-layers.

\newpage
{
\begin{table}
\begin{tabular}{|c|c|c|c|c|c|c|c|} \hline
Phonon  &  $\omega - \omega_{0}$ & $\sigma_{H}$ & $a_{\lambda}$ &
 $b_{\lambda}$ & $u_{xx}$  & $u_{zz}$\\
 Symmetry  & (cm$^{-1}$) & GPa
 & cm$^{-1}$  & cm$^{-1}$ &  $\times$ (10$^{-3})$ & $\times$ (10$^{-3}$)\\ \hline
$E_{2}$(high) & 5 $\pm$1.5 & 1.20 $\pm$0.36 & -81 $\pm$1 &
-2486$\pm$1& 1.83 $\pm$0.55 & -2.13 $\pm$0.64 \\
\hline
\end{tabular}

\caption{Hydrostatic stress, phonon deformation potential
constants and strain tensor components of Ga$^{++}$ implanted GaN
epi-layers.}
\end{table}}

\end{document}